\pdfoutput=1
\documentclass[12pt]{iopart} 

\usepackage{graphicx}
\usepackage{wrapfig}
\usepackage{comment}
\usepackage[caption=false]{subfig}
\usepackage{longtable}
\usepackage{appendix}
\usepackage{hyperref}
\usepackage{setspace}
\usepackage{comment}

\usepackage{array}
\usepackage{siunitx}
\usepackage[sc]{mathpazo}
\usepackage[T1]{fontenc}

\newcommand{\tmcolor}[2]{{\color{#1}{#2}}}
\newcommand{\tmem}[1]{{\em #1\/}}
\newcommand{\tmop}{\text}
\newcommand{\tmrsup}[1]{\textsuperscript{#1}}
\newcommand{\tmtextbf}[1]{{\bfseries{#1}}}
\newcommand{\tmtextit}[1]{{\itshape{#1}}}

\begin{document}

\title{Al/Ti/Al phonon-mediated KIDs for UV-VIS light detection over large areas}

\author{L~Cardani$^1$, N~Casali$^1$, A~Cruciani$^1$, H~le Sueur$^2$, M~Martinez$^{3,1}$, F~Bellini$^{3,1}$, M~Calvo$^4$, M~G~Castellano$^5$, I~Colantoni$^4$, C~Cosmelli$^{3,1}$, A~D'Addabbo$^{6}$, S~Di Domizio$^{7,8}$, J~Goupy$^{4}$, L~Minutolo$^{3,1}$, A~Monfardini$^4$, M~Vignati$^1$}

\address{$^1$INFN - Sezione di Roma, Piazzale Aldo Moro 2, 00185, Roma -
Italy}
\address{$^2$CSNSM, Univ. Paris-Sud, CNRS/IN2P3, Universite' Paris-Saclay,
91405 Orsay, France}
\address{$^3$Dipartimento di Fisica - Sapienza Universit{\`a} di Roma,
Piazzale Aldo Moro 2, 00185, Roma - Italy}
\address{$^4$Institut Neel, CNRS/UJF, 25 rue des Martyrs, BP 166, 38042
Grenoble, France}
\address{$^5$Istituto di Fotonica e Nanotecnologie - CNR, Via Cineto Romano
42, 00156, Roma - Italy}

\address{$^6$INFN - Laboratori Nazionali del Gran Sasso, Assergi (L'Aquila)
I-67010 -- Italy}
\address{$^7$Dipartimento di Fisica - Universit{\`a} degli Studi di Genova,
Via Dodecaneso 33, 16146, Genova - Italy}
\address{$^8$INFN - Sezione di Genova, Via Dodecaneso 33, 16146, Genova -
Italy}
\ead{angelo.cruciani@roma1.infn.it}

\begin{abstract}
The development of wide-area cryogenic light detectors with baseline energy resolution lower than 20 eV RMS is essential for next generation bolometric experiments searching for rare interactions. Indeed the simultaneous readout of the light and heat signals will enable background suppression through particle identification.

Because of their excellent intrinsic energy resolution, as well as their well-established reproducibility, Kinetic Inductance Detectors (KIDs) are good candidates for the development of next generation light detectors. The CALDER project is investigating the potential of phonon-mediated KIDs.

The first phase of the project allowed to reach a baseline resolution of 80 eV using a single KID made of aluminium on a 2x2 cm\tmrsup{$2$} silicon substrate acting as photon absorber. In this paper we present a new prototype detector implementing a trilayer aluminium-titanium-aluminium KID. Taking advantage of superconducting proximity effect the baseline resolution improves down to 26 eV.
\end{abstract}

{\maketitle}

\section{Introduction}\label{sec:introduction}

In the last decade, Kinetic Inductance Detectors
(KIDs){\cite{day}} underwent a rapid development, allowing their successful
application to millimeter {\cite{monfardini}} and UV astronomy
{\cite{mazin}}. Different projects are currently on-going with the aim to
apply this technology to other fields, ranging from the search
of extrasolar planets \cite{mazin2} to X-ray spectroscopy
{\cite{quaranta,faverzani}}.

The CALDER project {\cite{vignati}} is developing large area phonon-mediated
KIDs with the aim of detecting small amounts of ultraviolet-visible (UV-VIS) light. Such detectors will play a
crucial role for the next-generation bolometric experiments searching for
neutrinoless double-$\beta$ decay. CUORE {\cite{cuore}}, the current leading bolometric experiment, is an array of 988 TeO$_{2}$ crystals read by NTD sensors.
Its sensitivity to double-$\beta$ decay is mainly limited by the background induced by the $\alpha$ radioactivity of the material surrounding the detector at cryogenic temperature, mainly copper. The possibility to distinguish different particles would allow to overcome this limit. Particle identification could be accomplished by measuring the Cherenkov light ($\approx$ 100 eV) emitted in a TeO$_{2}$ crystal by electrons, that are the signal candidates, but not by $\alpha$ particles {\cite{tabarelli,casali2,casali3}}.
CUPID\cite{Wang:2015raa,invhie}, the planned	 upgrade of CUORE, will therefore require light detectors
able to monitor the whole face of a crystal ($\approx$ 25 cm\tmrsup{$2$}) with
a baseline energy resolution good enough to clearly detect this tiny amount of
light (< 20 eV RMS).

One of the limits of KIDs is that they can reach a maximum sensitive area of a few mm$^2$\cite{zmu}. To increase the sensible area to several cm$^{2}$ we opted for a phonon-mediated approach\cite{swenson}: photons are absorbed by the silicon substrate on which the KID is deposited and produce athermal or ballistic phonons able to reach the KID, where they break Cooper pairs and generate a signal.
The use of KIDs as phonon-mediated detectors was proposed for the detection of Dark matter \cite{golwala}, cosmic rays {\cite{swenson}} and X-rays {\cite{cruciani,moore}}.
The application of such detectors to detect visible light was initially studied by the CALDER project, using an array of 4 KIDs
{\cite{cardani}}, made of aluminium and lithographed on a $2 \times 2$cm$^{2}$, 300 $\mu$m thick silicon substrate. Recently it was demonstrated a baseline resolution of 82 eV using a single KID made of aluminium and combining both phase and amplitude readout, that had respectively a baseline resolution of 105 eV and 115 eV\cite{cardani2}. In the following we describe a resolution improvement down to 26 eV, obtained using a trilayer KID made of titanium (Ti) and aluminium (Al).

\section{Detector description}\label{sec:Detector}

KIDs base their working principle on the kinetic inductance of Cooper pairs,  which can be modified by an energy release able to break them into quasi-particles (QPs). If the superconductor is inserted in a resonant RLC circuit with high quality factor ($Q>10^{3}$), the density variation of QPs modifies the transfer function $S_{21}$ of the resonator both in phase and in amplitude. 
The phase readout is usually preferred since the phase response is larger (up to a factor 10) and is given by {\cite{mazinphd}}:
\begin{equation}
\label{eq1}
  \frac{d \phi}{dE} = \eta \frac{\alpha S_{\phi} (\omega, T) Q}{N_0 V
  \Delta_0^2}
\end{equation}
where $\eta$ is the energy to QPs conversion
efficiency, $\alpha$ is the fraction of kinetic inductance with respect to total inductance, $\Delta_0$
is the superconducting gap, $N_0$ is the single spin density of states
and $V$ the detector volume. $S_{\phi} (\omega, T)$ is a dimensionless factor given by the Mattis-Bardeen theory, describing the phase responsivity as a function of 
frequency and effective temperature \cite{mattis}. Finally $Q$ is the quality factor of the resonator. The detector described in \cite{cardani2}, made of a 60 nm thick Al film, featured $\eta=7.4 \%$, $\alpha=2.5\%$, $Q=1.5 \times 10^5$, $\Delta_{0}=179\ \mu$eV, $V=2.4 \times 10^{5}\ \mu$m$^{3}$, $S_{\phi}=2.85$, while for aluminium $N_0=1.72 \times 10^{10}$ eV$^{-1}\mu$m$^{-3}$.

One way to improve the responsivity is to increase the fraction of kinetic inductance $\alpha$ and to reduce the
superconducting gap $\Delta_0$, without affecting significantly the other parameters in Eq.
\ref{eq1}. This can be achieved by exploiting the
superconducting proximity effect between aluminium and a
lower gap superconductor, such as titanium. Titanium also features a higher kinetic inductance due to its higher penetration depth \cite{kittel} leading to a significant increase of $\alpha /
\Delta^2_0$. However titanium was never demonstrated to work as
a high quality factor resonator, possibly because of its high sensitivity to
contaminants {\cite{impu}}, yielding dissipative conducting states.
This leads us naturally to protect titanium with aluminium, which is known to have an almost
lossless self-protecting oxide layer.  Moreover, to efficiently collect phonons from the substrate into the
superconductor it is important to have an acoustic impedance matching between
the substrate and the KID. In that respect, aluminium provides a better coupling to silicon than
titanium \cite{match}. The geometry, we present, is thus an Al-Ti-Al trilayer with the following thicknesses: Al 14 nm / Ti 33 nm / Al 30 nm.

The fabrication of the chip was made according to the following steps. The intrinsic 380 $\mu$m thick Si <100> wafer is dipped
for 1 min in dilute 5\% HF solution and then rinsed in deionized water water for 1 min. The
wafer is then moved to an electron beam evaporator
dedicated to superconducting materials, where the layers are subsequently
evaporated at rates of 0.5 nm/s in a residual vacuum of $\tmop{$4 \times 10^{-8}$ mb}$. This is important especially for titanium
to avoid trapping impurities in the layer {\cite{impu}}.
The film is patterned using standard UV lithography followed by a
two-step wet etching. In a first step, the aluminium is removed
using a commercial etchant (H$_3$PO$_4$/HNO$_3$/CH$_3$COOH). Then, in the second step, the titanium
and the bottom aluminium layers are etched in diluted HF
(0.05\%) at a rate of about 0.2 nm/s.

\begin{figure}[t]
  \begin{center} 
	\includegraphics[width=.58\textwidth]{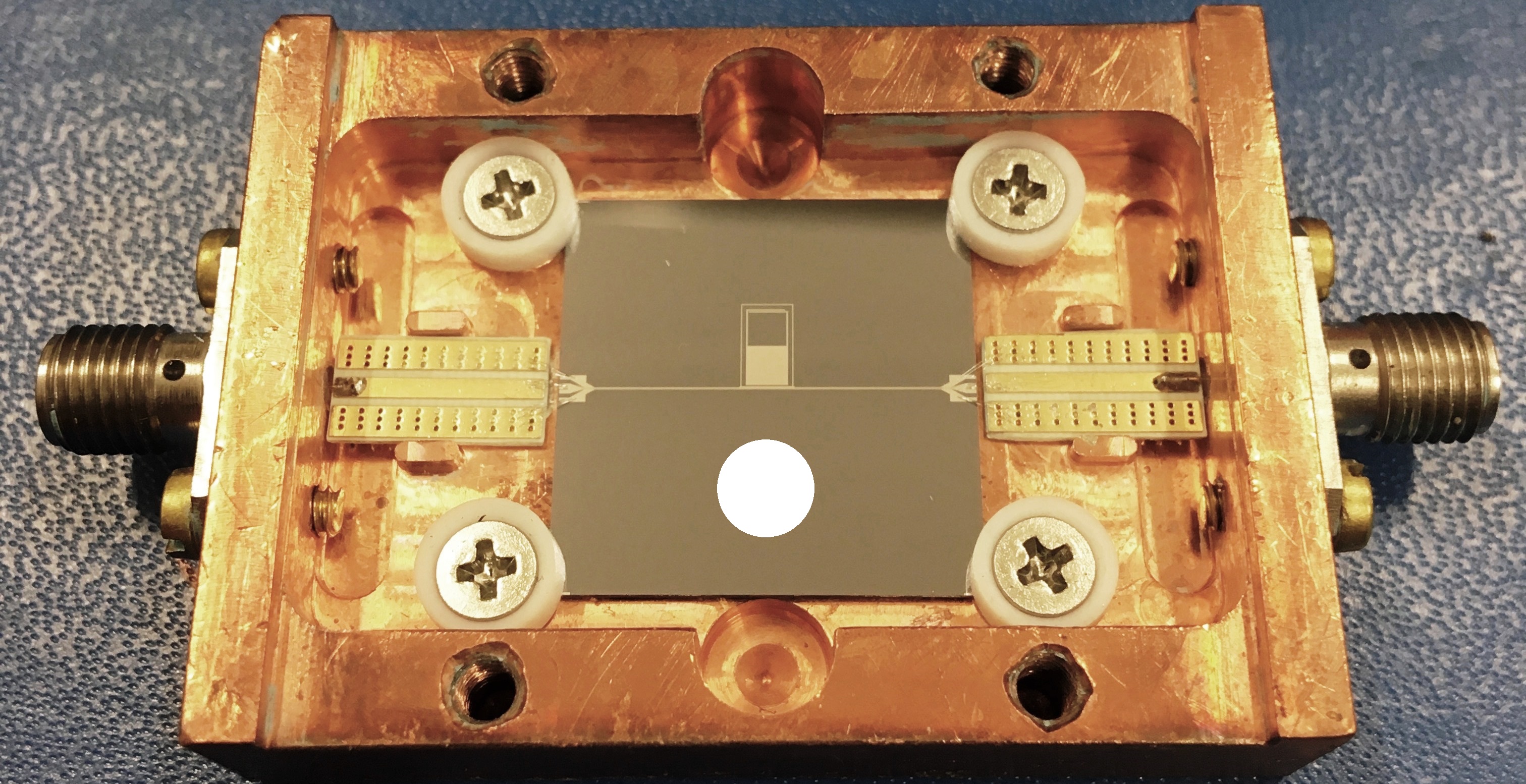} 
	\end{center} 
	\caption{\label{sample}The detector, made by a KID on a 2x2 cm\tmrsup{$2$} silicon substrate, is assembled in a copper structure through 4 PTFE supports. The circle show the position and size of the spot of the optical system used to illuminate the substrate.} 
\end{figure}

The sensor implements a LEKID design {\cite{doyle}} identical to the Al detector
in {\cite{cardani2}}. Its active area consists of an inductive
meander made of 30 connected strips of
62.5{\hspace{0.17em}}$\mu$m$\times$2{\hspace{0.17em}}mm. The meander is closed
with a capacitor made of 2 interdigitated fingers of
1.2{\hspace{0.17em}}mm$\times$50{\hspace{0.17em}}$\mu$m. The silicon substrate
is diced into a 2x2 cm\tmrsup{$2$} detector with the KID in the center as shown in Figure \ref{sample}.
The chip is assembled in a copper structure and held by 4 PTFE supports with total
contact area of about 3{\hspace{0.17em}}mm\tmrsup{$2$}. The copper holder is thermally anchored to the mixing chamber of a \tmrsup{$3$}He/\tmrsup{$4$}He dilution refrigerator with base
temperature lower than 10{\hspace{0.17em}}mK. 
\section{Electrical and thermal characterization of the detector}\label{sec:analysis0}

\begin{figure}[t]
  \begin{center}
	\includegraphics[width=.70\textwidth]{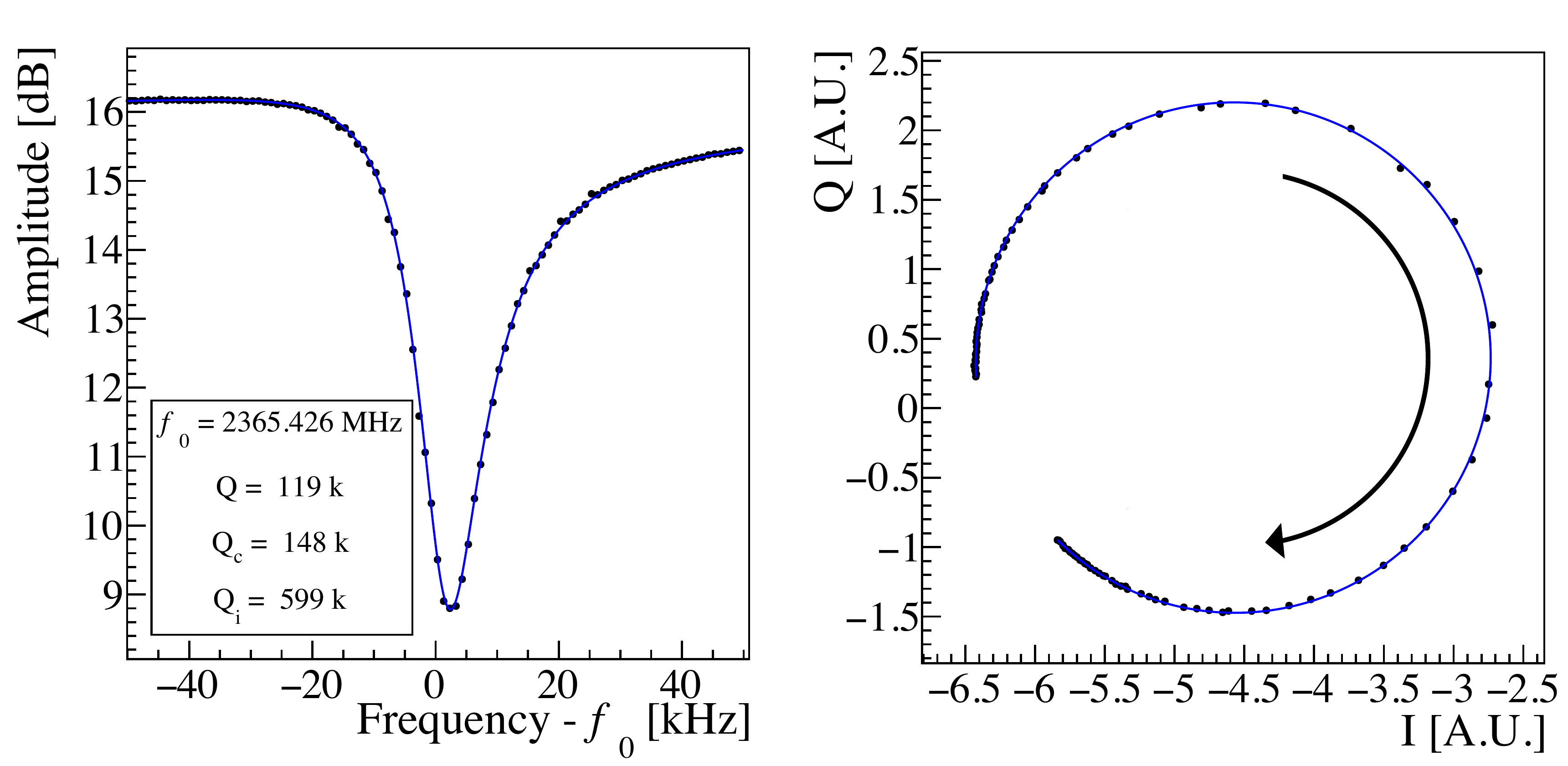}
\end{center}
\caption{\label{fig:resonances} Complex transmission
  coefficient ($S_{21}$) around resonance compared to fit (blue line). Left: $|S_{21}|$ vs frequency. Right: Same data in the complex IQ plane. The arrow indicates indicates the direction of increasing frequency.}
\end{figure}

The electrical characterization of the KID is based on the study of the
complex transmission ($\tmcolor{red}{S_{21}}$) of the resonator around the
resonant frequency (Figure~\ref{fig:resonances}). The frequency dependence of the
transmission of a resonator is ideally well
described by a single-pole approximation. It is however
necessary to take into account also impedance mismatches, power distortion and
electronics non-linearity to perform a rigorous estimation {\cite{casali}}. This measurement allows
\tmcolor{red}{one} to extract the coupling quality factor $Q_c$, the internal
quality factor $Q_i$ \tmcolor{red}{and the total quality factor $Q = (Q^{-
1}_c + Q_i^{- 1})^{- 1}$}.

The resonant frequency is found to be close to 2.4 GHz. $Q_c$ is $1.48 \times 10^5$ in very good agreement with the value predicted by the SONNET electromagnetic simulation of about $1.5 \times 10^5$. $Q_i$ is estimated as $6 \times 10^5$, lower
than our typical samples made of aluminium \cite{cardani2},
but high enough to allow the detector to operate well.

The critical temperature and the resistance of the trilayer
film have been measured in a 4-terminal geometry. The superconducting transition was
measured at (805 $\pm$ 10) mK with a normal state resistance per
square $R_{sq} = 0.58 \ \Omega$/sq above the transition. We estimate theoretically the $T_C$ of the trilayer following \cite{Catalano15}, assuming perfect
interface transparency to phonons across the layers and a superconductor total thickness
comparable or lower to its coherence length. Solving the Usadel equations \cite{usadel} within these
assumptions yields an analytical dependence of $T_C$ on the thickness of
the layers, similar to that found in the Cooper model \cite{cooper}.
To apply straightforwardly this bilayer model to the Al/Ti/Al trilayer, we treat it as a bilayer Al-Ti taking the average thicknesses of the Al
layers \cite{lesueur}. The expected critical temperature from this model is $T_{C} = 720\ \tmop{mK}$, assuming values of $T_{C}$ for the single layers that take into account the increase of critical temperature in thin films, as described in \cite{strongin}. The discrepancy between the expected $T_C$ and the measured value may come from the limitations of the model: we do not know the precise values of the electron-phonon coupling in all the layers, the actual transparency of
the interfaces and also the actual $T_{C}$ for each layer. A more complete model was recently proposed in \cite{zhao} and its experimental validation will be the subject of a separate work. 

Using the SONNET electromagnetic simulations
software we estimate the amount of kinetic inductance $L_k$ needed to shift the resonant frequency from the value simulated for a perfect conductor ($L_k=0$) to the measured value. The simulation is calibrated against a detector with identical geometry but in aluminium.  $L_k$  of the calibration device was calculated independently using direct measurements of $T_C$,$R_sq$, film thickness and an estimation of $\alpha$ from resonant frequency shift with temperature.
 We obtain a value of 1.4 pH/sq, which is significantly higher than the value of 1.0 pH/sq obtained using the BCS form described in \cite{zmu}, which depends on $T_C$ and $R_{sq}$. This is a first hint that our case of inhomogenous superconductor cannot be well described by the BCS theory.
The obtained value of kinetic inductance translates into a kinetic inductance fraction $\alpha$=17\%, much larger than that obtained in aluminium
($\alpha$=2.5\%) using the same design {\cite{cardani2}}.

\begin{figure}[t]
  \begin{center} 
	\includegraphics[width=.75\textwidth]{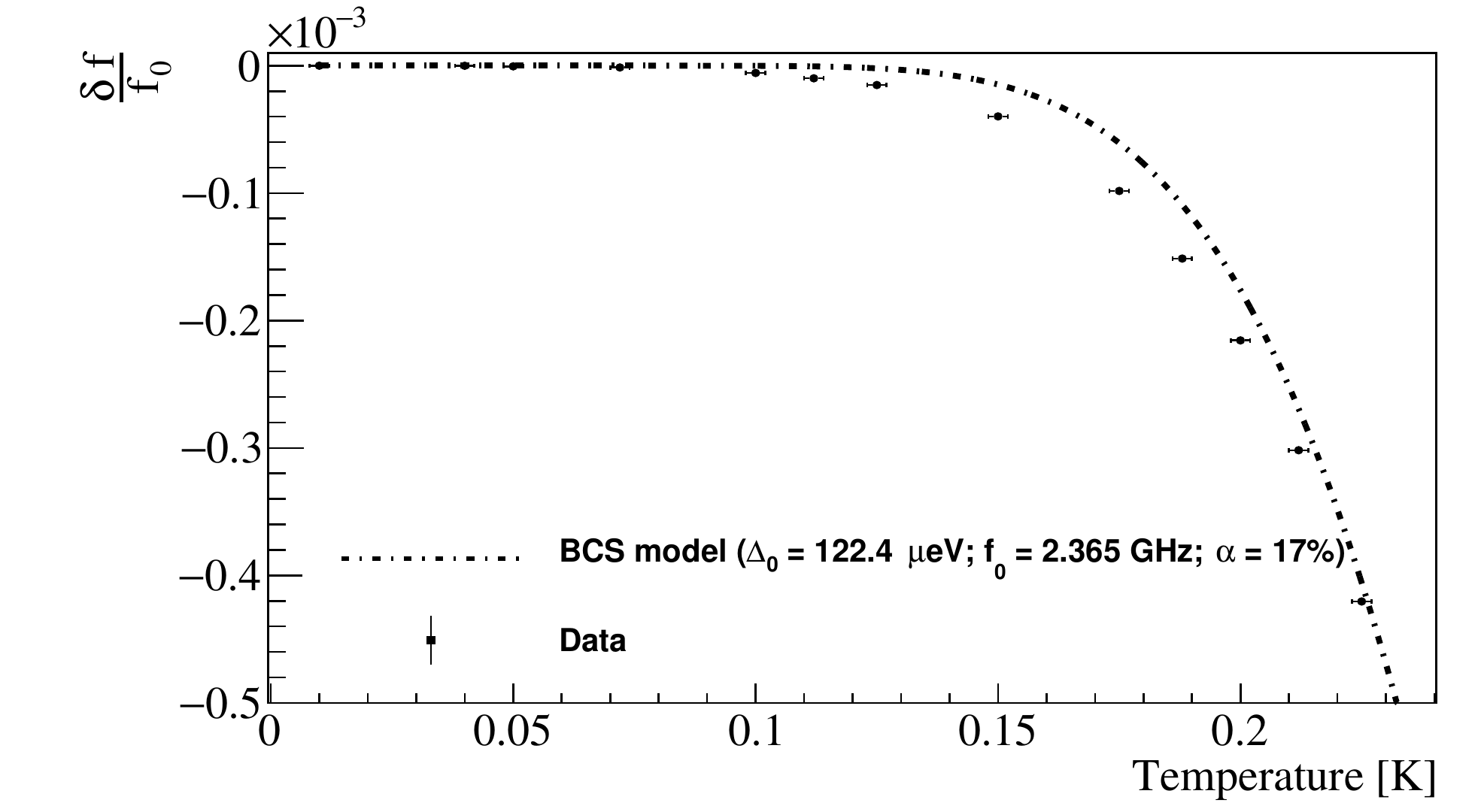}
  \end{center}
	\caption{\label{fig:alpha} Measured relative variations of
  the resonant frequency $(f - f_0) / f_0$ as a function of temperature (black
  squares), compared to the Mattis-Bardeen theory (dashed
  lines)} 
\end{figure}

We have also measured the relative variations of the
resonant frequency $(f - f_0) / f_0$ as a function of the temperature (Figure
\ref{fig:alpha}). These data are compared with a model developed
by Gao {\tmem{et al.}}{\cite{Gao}} using the Mattis-Bardeen theory. The model is calculated for our best estimation of the parameters ($T_C = 805$ mK and $\alpha =
17\%$). Again the BCS form cannot describe properly the case of our
inhomogeneous superconductor and we can indeed see the data departing from
the BCS model at intermediate temperatures. The interpretation of these measurements needs a dedicated model, whose development is beyond the scope of this paper. 

\section{Detector performance}

Our experimental setup allows to back-illuminate the detector using two different sources: a non-collimated X-ray $^{55}$Fe
source and an optical fiber coupled to a fast room-temperature LED.
The $^{55}$Fe source illuminates in an almost uniform way the whole substrate with a rate of 1 Hz.
The LED emits photons at 400{\hspace{0.17em}}nm in the typical range of Cherenkov light. Depending on the LED
pulse duration and/or amplitude, a variable number of photons can be sent onto the substrate. The whole optical system is intially calibrated with an accuracy of 10 \% using a PMT at room temperature and then cross-calibrated with the $^{55}$Fe source. The position and size of the optical system spot on the substrate is shown in Figure \ref{sample}.

The resonator is excited and probed with a monochromatic tone at the resonant frequency $f_0
$. The output signal is fed into a CITLF3 SiGe cryogenic low noise amplifier, downconverted at room temperature using a superheterodyne electronics and then digitized
with an acquisition card at a sampling frequency of 500{\hspace{0.17em}} kSPS.
Time traces of up to 12{\hspace{0.17em}}ms of the real (I) and imaginary (Q) parts of $S_{21}$ are acquired following a software trigger. Finally I and Q variations are converted into changes of phase
and amplitude relative to the center of the resonance loop. A detailed description of the experimental setup of our laboratory at INFN
Rome, including the room-temperature electronics and the acquisition software, can be found in references \cite{vignati,bourrion}.

As already mentioned, the phase response is typically larger than the amplitude one. However the phase noise exhibits often an excess noise, which sometimes makes it advantageous to consider both phase and amplitude responses in the analysis of the detector sensitivity. When amplitude and phase noises are dominated by the noise of the low noise amplifier, they decrease with microwave power as $S_{\varphi}
\propto P_{\tmop{in}}^{- 1 / 2}$; however at large powers the
non-linearities of the kinetic inductance \cite{swensonjap} and heating effects \cite{pippo} become relevant and can suppress the signal. The
recombination time {$\tau_{QP}$, which determines the signal integration time \cite{cruciani2}, is also reduced by heating effects, lowering the signal-to-noise ratio. For these reasons, it is important to
determine experimentally the bias power leading to the optimal signal-to-noise ratio.

\begin{figure}[t]
  \begin{center}
	\includegraphics[width=.65\textwidth]{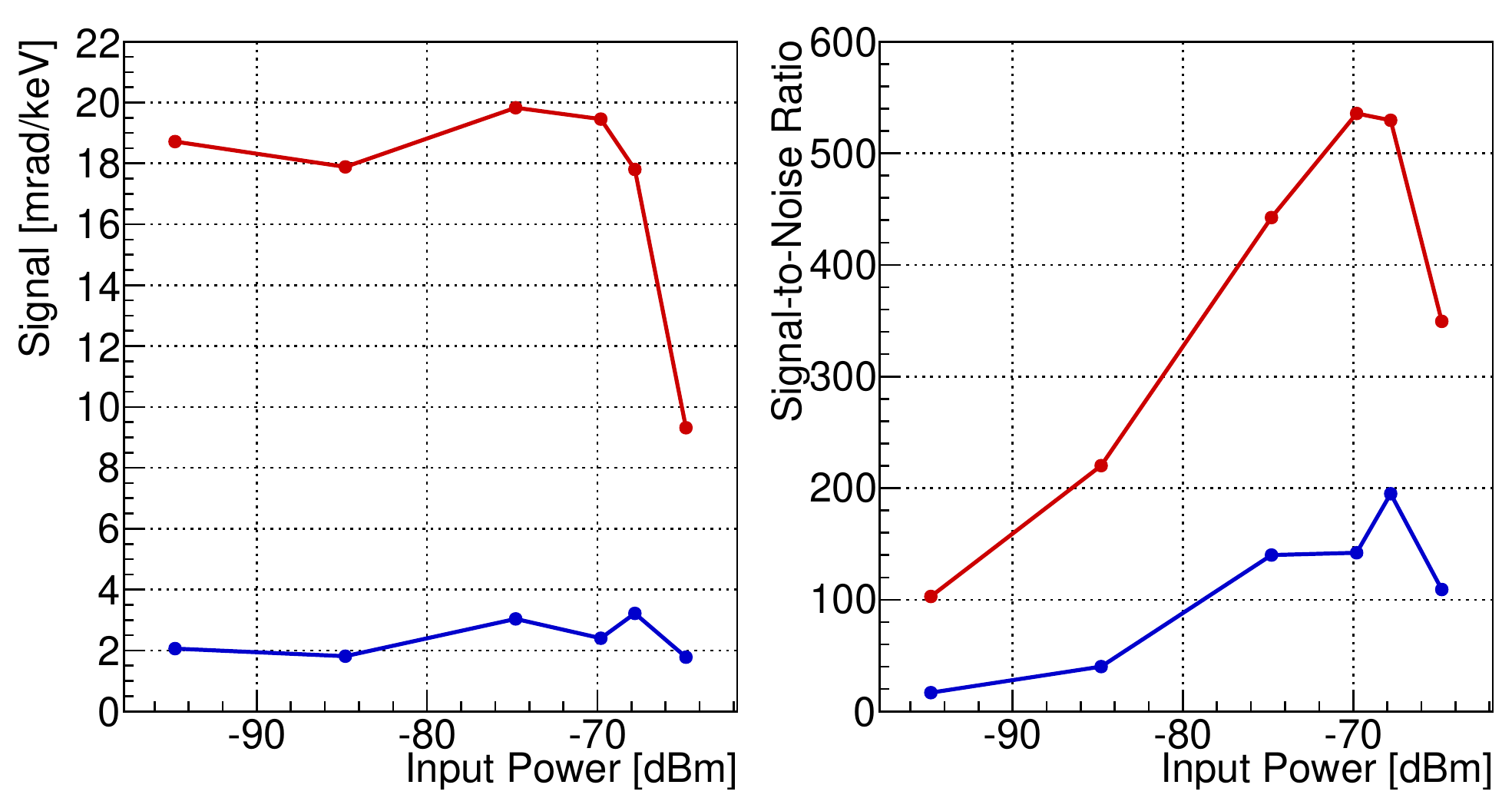}
\end{center}
  \caption{\label{fig:Psweep1} Results of the power scan, obtained by sending light pulses of 15.5 keV on the detector. Left: Phase (red) and amplitude (blue) signal 
height in mrad as function of the bias power. Right: Signal-to-Noise ratio as function of bias power. The optimal working point is chosen at -70 dBm. }
\end{figure}
For this purpose, we perform a microwave power scan from
about $-95${\hspace{0.17em}}dBm to $-65${\hspace{0.17em}}dBm at the
input of the sample, while the monitoring signal and noise levels,
as well as the pulse shape, in both the phase and amplitude readouts. A reference signal is produced by sending light pulses of 15.5 keV. The signals are processed using a matched filter, a software algorithm that allows to improve the signal-to-noise ratio by suppressing the signal frequencies that are mostly affected by noise \cite{gatti}. The height of the filtered pulses is reported in Figure \ref{fig:Psweep1} left. The phase
response is quite flat with increasing power around 19 mrad/keV, a sizeable improvement with respect to the 6
mrad/keV obtained with the Al detector in \cite{cardani2}. A similar increase is observed for the amplitude response, around 2.5 mrad/keV with respect to the 0.6 mrad/keV obtained with the Al detector. 

In Figure \ref{fig:Psweep1} right we show the signal-to-noise ratio, obtained by dividing the amplitude of the pulses by the RMS of their baseline (after the matched filter). We determine an optimal working power of $- 70 \tmop{dBm}$, about 10dB lower than what obtained with
aluminium. The efficiency $\eta$ at the optimal working power is estimated as in \cite{cardani2} and amounts to $\eta=(7.5 \pm 1) \%$ consistent with what found in \cite{cardani2}.
\begin{figure}[t]
\begin{center} 
\includegraphics[width=.65\textwidth]{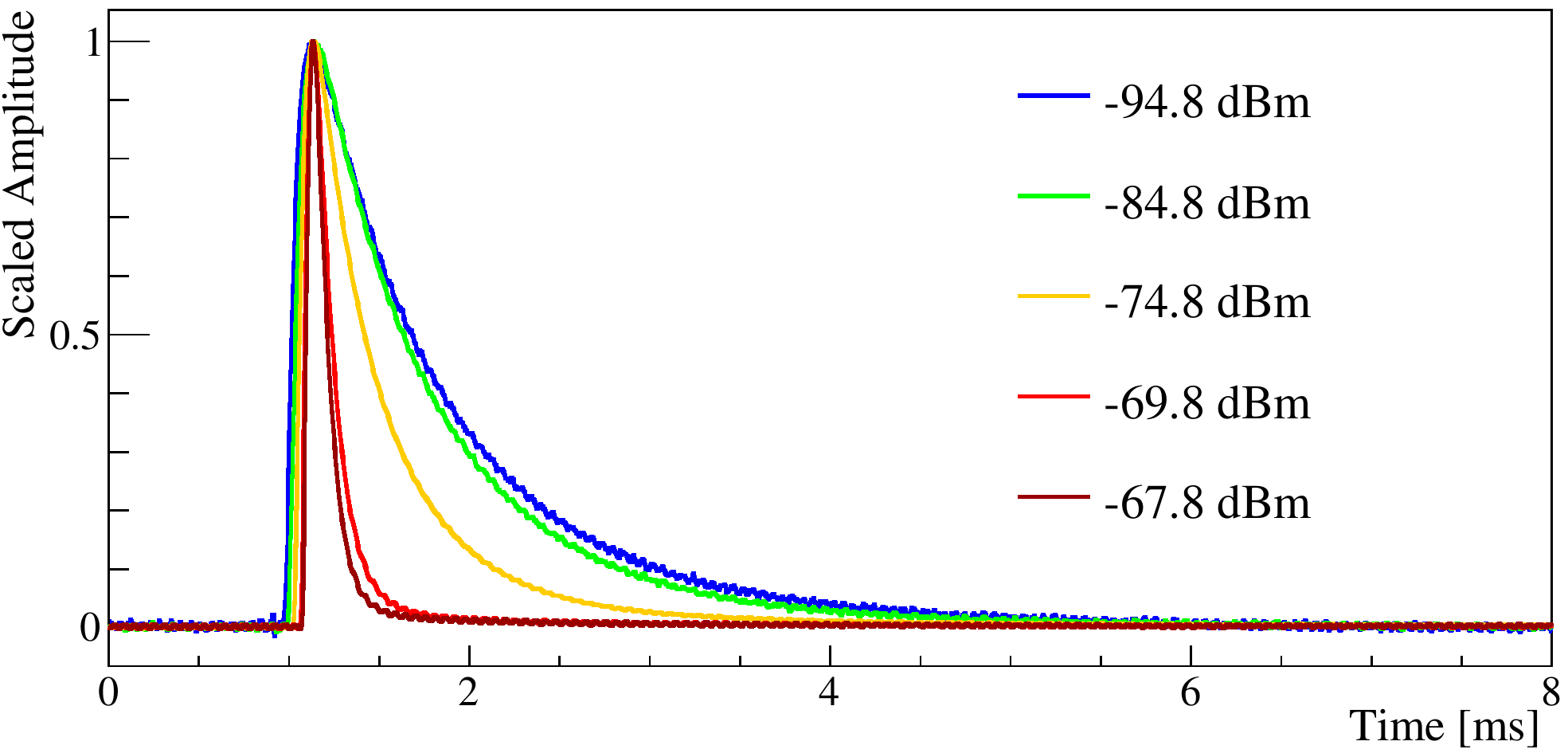}
\end{center} 
\caption{\label{fig:tau} Response of the resonator to light pulses of 15.5 keV for increasing microwave powers. The time traces are obtained by averaging many waveform to reduce the random noise and their amplitude is normalized to 1 to highlight the shape variation.}
\end{figure}


The trailing edge of the pulse is well described by a single exponential law with a decay time that varies between 800 $\mu$s and 100 $\mu$s with increasing microwave power (Figure \ref{fig:tau}). As shown in \cite{cruciani2}, this decay constant can be interpreted as $\tau_{QP}$. The noise power spectrum at optimal power is reported in Figure~\ref{fig:noise} for
both the phase  and amplitude readouts. The flat noise observed in the amplitude readout
and in the high frequency region of the phase readout, is consistent with the
noise temperature of the cold amplifier ($T_N \sim 6 \hspace{0.17em} K$). This noise
is exceeded by a low-frequency phase noise, whose spectral index is found to be $0.49 \pm 0.05$, consistent with the presence of a two-level system noise in the detector, observed often in KIDs \cite{gao_TLS}. In order to compare the obtained TLS noise with other works, we convert the noise in fractional frequency noise as in \cite{zmu} and get $S_{TLS}$(f=1 KHz)=$1 \times 10^{-21}$Hz$^{-1}$, that is among the best values reported in \cite{zmu}. Both power spectra show also peaks due to  the readout electronics. These peaks are strongly suppressed by the matched filter and the measured energy resolution is always found consistent with the predictions from the noise power spectrum and the matched filter.

\begin{figure}[t]
  \begin{center}
	\includegraphics[width=.72\textwidth]{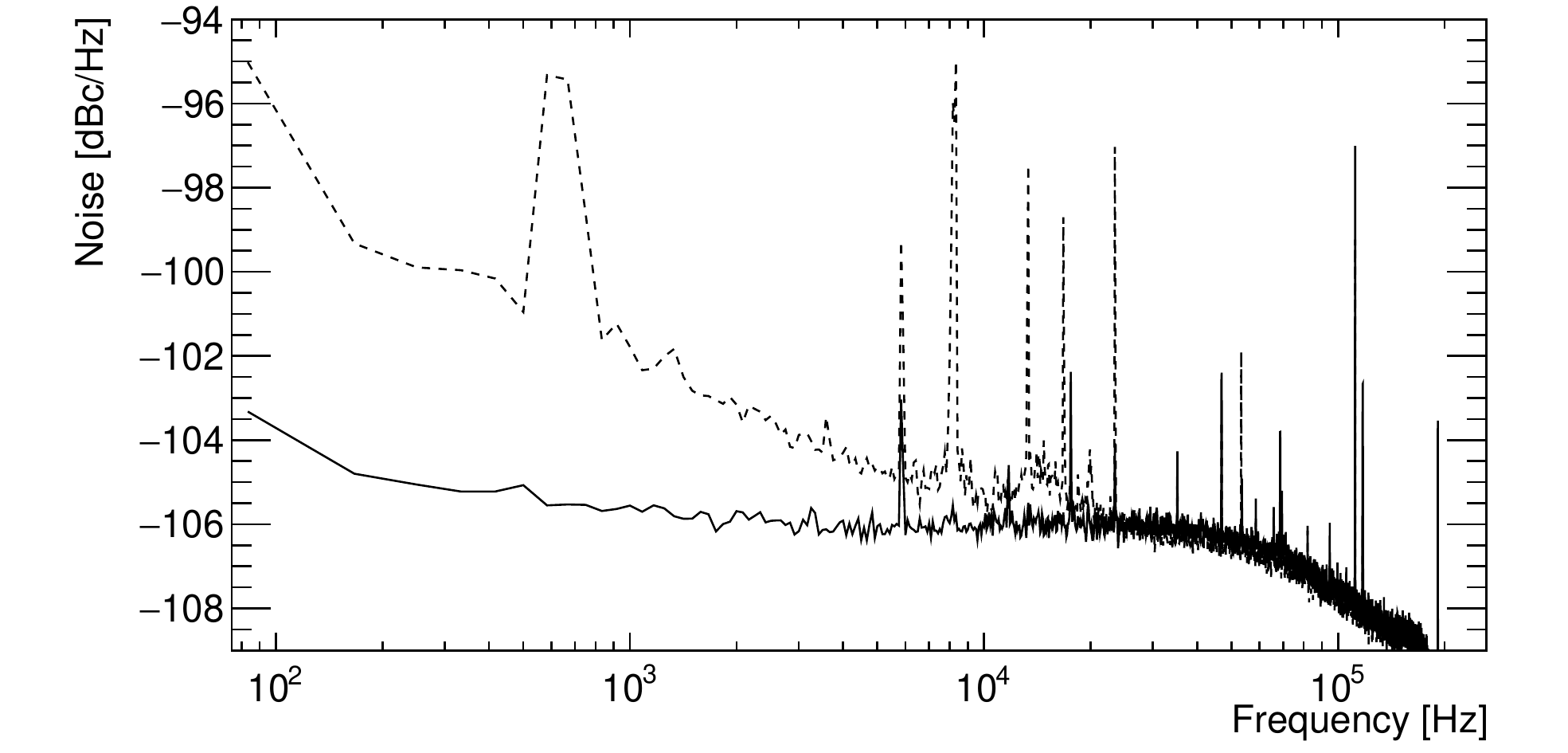}
  \end{center}
	\caption{\label{fig:noise}Average noise power spectrum in phase (dotted line) and amplitude (solid line) readouts. More details in the text.}
\end{figure}
\begin{figure}[b]
  \begin{center}
	\includegraphics[width=.48\textwidth]{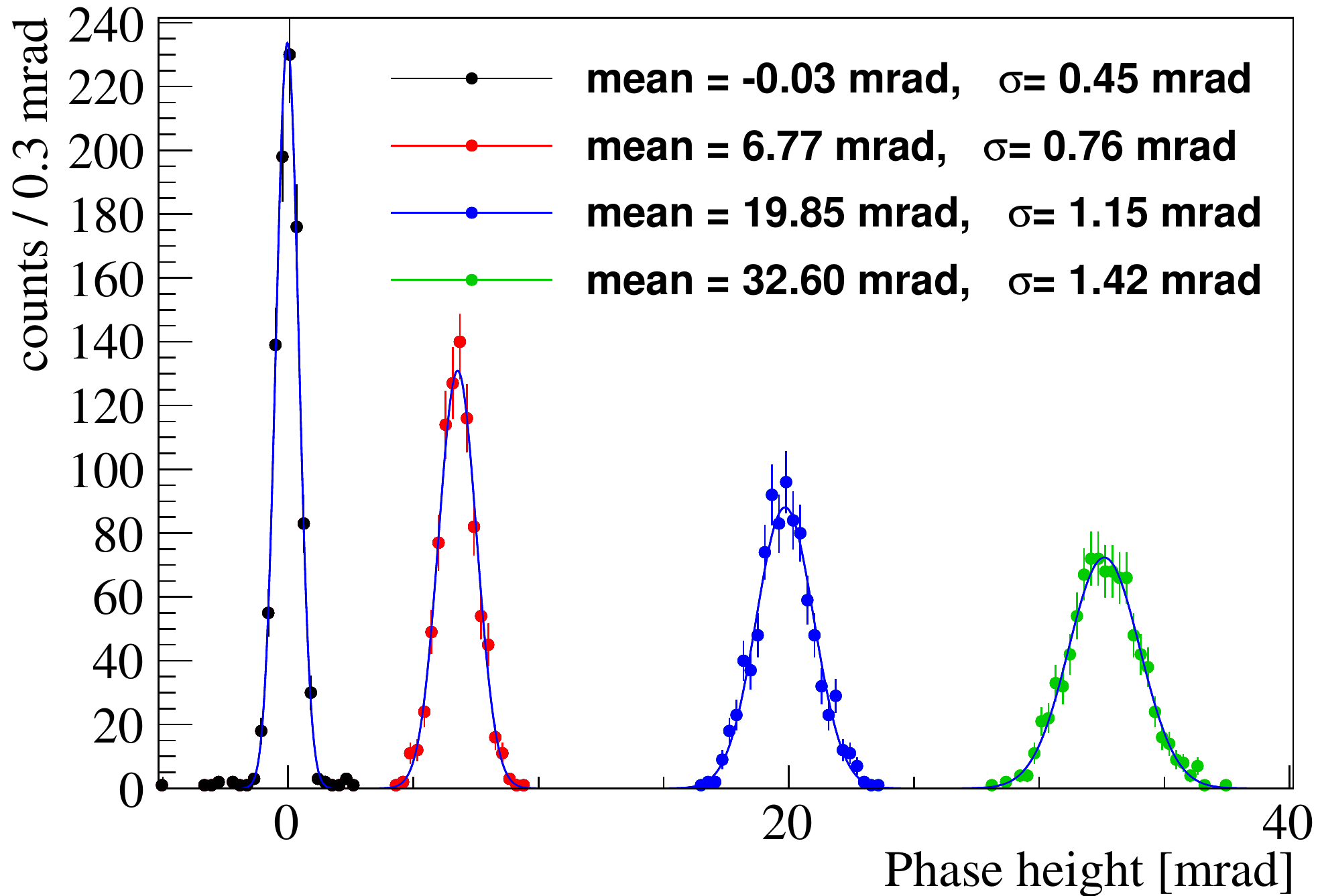} \quad \includegraphics[width=.48\textwidth]{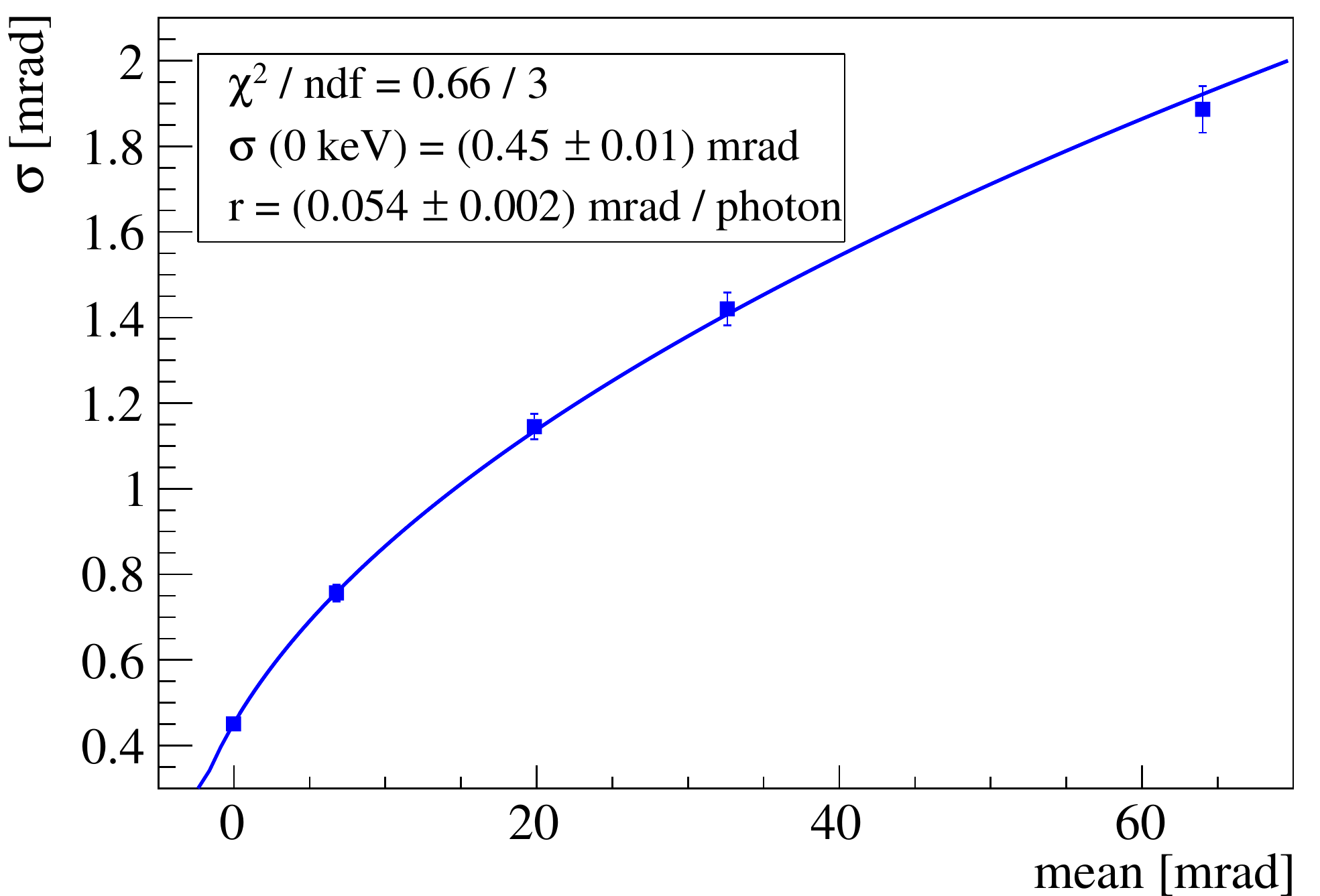}
	\end{center}
	\caption{\label{selfcalib}Left: Histograms of pulse heights in the phase readout for light signals of different energies. Right: Standard deviation $\sigma$ versus mean $m$ of the Gaussian distributions shown in the left panel. The plot includes one more point at 65 mrad not shown in the left panel to not compress the scale for the other distributions. The curve is well described by the Poisson statistics of photons. See the text for more details.}
\end{figure}

To obtain a solid estimation of the detector performance we implement a calibration based on the Poisson statistics of the photons absorbed into the substrate. The detector is biased at the optimal microwave power of -70 dBm and is illuminated with optical pulses of energies in the range 0 - 3 keV. For each energy the distribution of the pulse heights in the phase readout is found to be well described by a Gaussian with mean $m$ and standard deviation $\sigma$ (Figure \ref{selfcalib} left).
The $\sigma$ vs $m$ data are then fitted (Figure \ref{selfcalib} right) assuming that two uncorrelated components contribute to the detector resolution: the baseline resolution ($\sigma_0$) and the Poissonian term from the photon counting that we model as $\sqrt{m \times r}$\footnote{The number of photons absorbed in the detector is $N_{ph}=m / r$. According to the Poisson statistics one has $\sigma_{ph}=\sqrt{N_{ph}} \times r =\sqrt{m/r}\times r=\sqrt{m \times r}$.}, where $r$ is the responsivity per photon which is estimated from the fit: $\sigma = \sqrt{\sigma_0^{2} + m\times r}$. By dividing $r$ by the energy of a single 400 nm photon we
obtain a responsivity of ($17.54 \pm 0.58$ mrad/keV) in good
agreement with the calibration against $^{55}$Fe presented previously in this Section. Using this energy calibration we find a phase readout baseline resolution $\sigma_{\phi} = (25.64 \pm 0.85)$ eV. Applying the same procedure to the amplitude readout, we obtain $\sigma_A = (77.3 \pm 2.5)$ eV.

\section{Conclusions and discussions}

In this paper we presented the results obtained using an Al/Ti/Al trilayer film
to develop a phonon-mediated KID for the detection of UV-VIS light on large surfaces.
The sample features $T_C = 805$mK, $Q_i \simeq$600 k
and a low-power recombination time $\tau_{QP}=800\mu$s. The main features of the detector are compared in Table 1 with the Al detector, presented in \cite{cardani2}. The lower critical temperature increases significantly the fraction of kinetic inductance $\alpha$ and thus the responsivity of the detector (factor 3 in the phase readout). The integrated phase noise after the matched filter improves of about 25 \%. This improvement can be ascribed likely  to the change in the fabrication process: the new trilayer detector was made by etching while the previous detectors were made by lift-off. The overall result is the improvement of the RMS baseline resolution for phase readout from 105 eV down to 26 eV. The amplitude responsivity also improves (factor 4). Since the amplitude noise is dominated by the LNA noise, the lower bias power implies however a worsening of the integrated noise by a factor 3. Therefore the amplitude baseline resolution improves only from 115 to 77 eV.
The obtained energy baseline resolution approaches the target resolution of the CALDER project of 20 eV. The last phase of the project -
currently in progress- consists in the development of light detectors of 5x5 cm\tmrsup{$2$}.

\begin{table}[t]
	\centering
		\begin{tabular}{|c|c|c|} 
		\hline
			& \textbf{Al KID} & \textbf{Al/Ti/Al KID}\\ \hline
			$T_c$ [mK] & 1.18 & 0.805 \\
			$Q_i$ [k] & > 2000 & 600 \\
			$\alpha$ & 0.025 & 0.17 \\
			Phase resp. [mrad/keV] & 5.8 & 17.6 \\
			Amplitude resp. [mrad/keV] & 0.6 & 2.5 \\
			Integ. phase noise [mrad] & 0.61 & 0.46 \\
			Integ. amplitude noise [mrad] & 0.069 & 0.190 \\
			$\sigma_\phi$ [eV] &105 & 26 \\
			$\sigma_A$ [eV]& 115 & 77 \\
			$\sigma_{T}$ [eV] & 82 & 26  \\
			\hline
		\end{tabular} 
			\caption{\label{table}Main features of the detector presented in this paper compared with the Al detector described in \cite{cardani2}. $\sigma_{T}$ is the resolution obtained for the combination of phase and amplitude readout, using a 2D matched filter \cite{cardani2}. }

\end{table}

\section*{Acknowledgements}

This work was supported by the European Research Council (FP7/2007-2013) under
contract CALDER no. 335359 and by the MIUR under the
FIRB contract no. RBFR1269SL. The authors thank the personnel of INFN Sezione
di Roma for the technical support, in particular M. Iannone, F. Pellegrino, L.
Recchia and D. Ruggeri. HLS, MC, JG and AM acknowledge support from the ANR
grant ELODIS2 no. ANR-16-CE30-0019-02. The authors thank also the anonymous reviewers for useful suggestions and comments.
\\

\end{document}